\documentclass[pageno]{jpaper}

\usepackage[normalem]{ulem}
\usepackage{graphicx}
\usepackage{listings}
\usepackage{xspace}

\usepackage{algorithm}
\usepackage[noend]{algpseudocode}

\usepackage{amsmath}
\usepackage{amsthm}
\usepackage{amsfonts}
\usepackage{array}
\usepackage{booktabs}
\usepackage{boxedminipage}
\usepackage{calrsfs}
\usepackage{caption}
\usepackage{color}
\usepackage{enumitem}
\usepackage{fancyhdr}
\usepackage{float}
\usepackage{xcolor}
\usepackage{hyperref}
\usepackage[capitalize,nameinlink]{cleveref}
\hypersetup{
	unicode=true,
        bookmarksnumbered=true,
	colorlinks=true,
	linkcolor={red!70!black},
	citecolor={red!70!black},
	urlcolor={blue!70!black},
	pdfborder={0 0 0}
}
\usepackage[capitalize,nameinlink]{cleveref}
\usepackage[T1]{fontenc}
\usepackage[utf8]{inputenc}
\usepackage{microtype}
\microtypecontext{spacing=nonfrench}
\usepackage{multirow}
\usepackage{nicefrac}
\usepackage{pifont}
\usepackage{sidecap}
\usepackage{url}
\usepackage{verbatim}
\usepackage{framed}
\usepackage{tcolorbox}
\usepackage{etoolbox}

\usepackage[backend=biber,style=acmnumeric,datamodel=acmdatamodel,autocite=plain,sortcites=true,url=false,maxbibnames=99,date=year]{biblatex}
\newif\ifshowcomment
\showcommenttrue

\ifshowcomment
	\newcommand{\todo}[1]{\textsf{\color{red}{[{TODO: #1}]}}}
	\newcommand{\pratyush}[1]{\textsf{\color{orange}{[P: {#1}]}}}
	\newcommand\inigo[1]{\textcolor{red}{IG: #1}}
    \newcommand{\esha}[1]{\textcolor{cyan}{EC: #1}}
    \newcommand{\chaojie}[1]{\textcolor{brown}{CZ: #1}}

\else
	\newcommand{\todo}[1]{}
	\newcommand{\pratyush}[1]{}
	\newcommand{\inigo}[1]{}
    \newcommand{\esha}[1]{}
    \newcommand{\chaojie}[1]{}
\fi

\usepackage{siunitx}
\newcommand{\papername}{POLCA\xspace}

\newcommand{\s}{\si{\second}\xspace}

\newcommand{\myparagraph}[1]{\vspace{\smallskipamount}\noindent\textbf{#1.\xspace}}

\newcommand{\myparagraphemph}[1]{\vspace{\smallskipamount}\noindent\emph{#1.\xspace}}
\newcommand{\eg}{\emph{e.g.}\xspace}

\newcommand{\ie}{\emph{i.e.}\xspace}

\begin{document}

\title{POLCA: Power Oversubscription in LLM Cloud Providers\vspace{-9pt}}
\author{\\
Pratyush Patel\thanks{Pratyush Patel is affiliated with the University of Washington, but was at Microsoft during this work.} ,
Esha Choukse,
\\
Chaojie Zhang,
\'{I}\~{n}igo Goiri,
Brijesh Warrier,
Nithish Mahalingam,
Ricardo Bianchini
\vspace{.2in}
\\
Microsoft Azure
}
\date{}
\maketitle

\thispagestyle{empty}
\begin{abstract}

Recent innovation in large language models (LLMs), and their myriad use-cases have rapidly driven up the compute capacity demand for datacenter GPUs. Several cloud providers and other enterprises have made substantial plans of growth in their datacenters to support these new workloads.
One of the key bottleneck resources in datacenters is power, and given the increasing model sizes of LLMs, they are becoming increasingly power intensive. In this paper, we show that there is a significant opportunity to oversubscribe power in LLM clusters. Power oversubscription improves the power efficiency of these datacenters, allowing more deployable servers per datacenter, and reduces the deployment time, since building new datacenters is slow.

We extensively characterize the power consumption patterns of a variety of LLMs and their configurations.
We identify the differences between the inference and training power consumption patterns.
Based on our analysis of these LLMs, we claim that the average and peak power utilization in LLM clusters for inference should not be very high.
Our deductions align with the data from production LLM clusters, revealing that inference workloads offer substantial headroom for power oversubscription.
However, the stringent set of telemetry and controls that GPUs offer in a virtualized environment, makes it challenging to have a reliable and robust power oversubscription mechanism.

We propose \papername, our framework for power oversubscription that is robust, reliable, and readily deployable for GPU clusters.
Using open-source models to replicate the power patterns observed in production, we simulate \papername and demonstrate that we can deploy 30\% more servers in the same GPU cluster for inference, with %
minimal performance loss. 

\end{abstract}

\section{Introduction}
\label{sec:introduction}

\begin{figure*}[ht] 
  \begin{minipage}[b]{0.44\textwidth}
    \centering
    \includegraphics[width=1\textwidth]{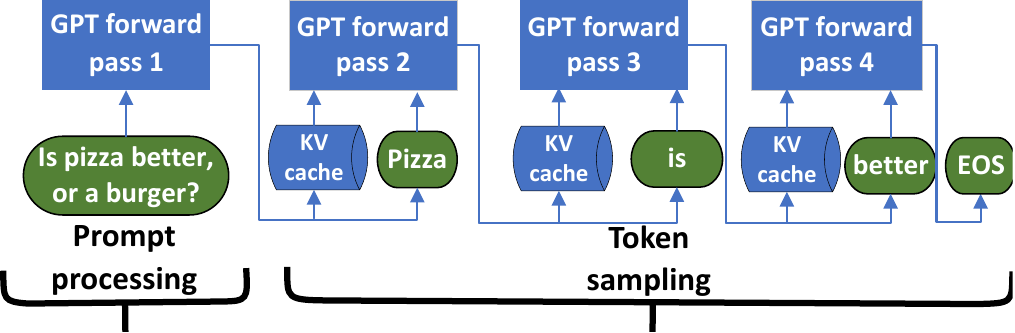}
    \caption{An example of the prompt and token phases in a GPT-style model.} 
    \label{fig:prompt_v_token}
  \end{minipage}%
  \hfill
  \begin{minipage}[b]{0.20\textwidth}
    \centering
    \includegraphics[width=0.8\textwidth]{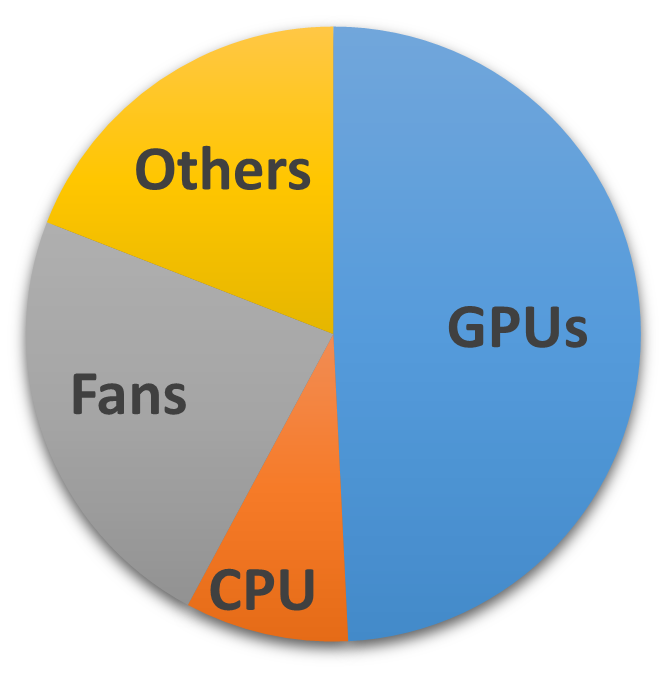} 
    \caption{Provisioned power ($8\times$A100-80GB server).} 
    \label{fig:A100Power}
  \end{minipage} 
  \hfill
  \begin{minipage}[b]{0.24\textwidth}
    \footnotesize
    \centering
    \begin{tabular}{cl}
    \toprule
    \textbf{Category}  & \multicolumn{1}{c}{\textbf{Models}}                  \\
    \midrule
    Encoder   & RoBERTa                \\
    Decoder   & GPT-NeoX-20B \\
            & OPT-30B* \\
            & BLOOM-176B* \\
    Encoder-Decoder & Flan-T5 XXL  \\
    \bottomrule
    \end{tabular}
    \caption{Our LLM workloads. *inference only.}
    \label{tab:workloads}
  \end{minipage}
\end{figure*}
\myparagraph{Motivation}
Datacenters and cloud providers today face a massive GPU capacity crunch due to the explosion in demand for large language models (LLMs)~\autocite{bender2021dangers}.
For example, OpenAI scaled up their clusters to 7,500 GPU servers to train LLMs like GPT-3~\autocite{openai_7500k8s}; Meta deployed an AI training supercluster with over 6,000 A100 GPUs~\autocite{meta_ai_supercluster}. 
This demand is only growing for training newer and larger models like Bard and GPT-4~\autocite{iiss_llm}.
The demand for inference is even larger than for training, and may constitute over 90\% of the overall LLM compute cycles~\autocite{patterson2022carbon,aws_inferentia_instances,tirias2019why}.
To keep up, several enterprises are making large investments into building new GPU clusters to run LLM workloads~\autocite{meta_ai_supercluster,openai_7500k8s}.
However, building new datacenters is expensive and carbon intensive~\autocite{iea_dc,luccioni2022estimating}; and crucially, doing so takes a long time which does not address the immediate demand.
Power, space, and cooling are the major bottlenecks in making datacenters denser.
Power-dense deployments like DGX A100s though, have power as their main bottleneck. Datacenters are deployed with a fixed power budget, based on generators and contracts with the utility companies~\autocite{fan2007power, smoothOperator, zhang_Flex, fu2011much}. Therefore, despite lower power utilization, without a proper power oversubscription mechanism, adding more GPU servers to an existing datacenter would push beyond the available power budget, and is therefore not an option. 

\myparagraph{Our work}
We extensively analyze the power consumption patterns and phases in LLMs. We investigate both, the training and inference serving, with various configurations representative of different LLM use-cases.
Based on this, it is clear that there are specific properties in the workloads that would lead to low power utilization at the inference cluster level, despite high power utilization at the server-level.
We then observe the power utilization of LLM inference and training at the cluster level in production, to verify our claims.
We observe that the power utilization does not peak to the level of the allocated power, despite individual servers peaking to their allocated power.
Instead, LLM inference clusters utilize only up to 80\% of the provisioned power, making them excellent candidates for power oversubscription.
In contrast, LLM training clusters offer a smaller headroom (about 10\%) since they incur in massive and coordinated power peaks due to large-scale synchronous training jobs.
In this work, our goal is to safely oversubscribe the provisioned power in existing and upcoming multi-tenant GPU clusters to reduce costs and address the capacity crunch of running LLM workloads.

Power oversubscription could sometimes result in power overload, and therefore cannot be deployed safely without implementing mitigation mechanisms.
This poses three main challenges for power oversubscription in GPU clusters.
First, since LLM workloads are GPU intensive, existing CPU-based power oversubscription techniques~\autocite{kumbhare2021prediction,li2020thunderbolt,wu2016_Dynamo,li2019scalable} are ineffective in these clusters. To deal with this, we explore the power distribution within the server to ascertain the impact the GPU can have on the server-level power.

Second, since LLMs are new and rapidly evolving, the efficacy of the throttling in reducing power usage, and the impact on application performance is not well understood.
We base our design on our characterization of the efficacy of GPU power capping and frequency scaling on modern LLM inference workloads.
We also target \emph{configurability} and \emph{robustness} to support the changing LLMs over time. 

And third, GPU power management poses its own set of challenges~\cite{patel2023towards}.
GPUs do not expose the plethora of well tailored power telemetry and control knobs that the CPU-based datacenters use to make cluster-level power throttling decisions.
Datacenters need out-of-band mechanisms to communicate with the devices in a controlled and time-sensitive way.
Although out-of-band GPU power management interfaces do exist, they are slow and unreliable, which complicate safe power throttling.
We address this with a double threshold solution, ensuring a safe time buffer before reaching peak cluster power utilization.
Our approach uses two cluster-level power thresholds for frequency throttling based on the inference priority.
If power usage remains insufficiently reduced, and we reach maximum cluster power utilization, all GPUs are rapidly throttled via the hardware powerbrake mechanism to prevent power outages.

Based on these insights, we design \papername, a robust, reliable power oversubscription framework for LLM inference clusters, which integrates with existing cluster-level power manager.
Using open-source models, we replicate power patterns from production LLM inference clusters for evaluation.
\papername boosts allocated server capacity by 30\% in existing inference clusters, with minimal power throttling events.
This improves power efficiency, reduces costs through fewer datacenters, and promptly meets the demand for running additional LLM workloads.

\myparagraph{Summary}
We make the following contributions:
\begin{itemize}[nosep,leftmargin=*]
    \item An extensive characterization of the power patterns of modern LLMs in their training and inference phases, with a deep dive into power consumption phases in inference with different configurations.%
    \item A characterization of the efficacy of existing GPU power management knobs, namely frequency scaling and power capping, at reclaiming power for LLM workloads.%
    \item  An overview of the power headroom available in production LLM clusters, that is in line with our characterization and understanding of these LLM workloads. 
    \item A case for building separate inference-optimized clusters, to increase the datacenter power utilization of LLM inference.
    \item \papername, a robust and reliable approach for modern GPU servers that enables safe and efficient power oversubscription in LLM clusters today, while meeting performance SLOs.
    \item An evaluation of \papername on a replication of power patterns in LLM inference production clusters. %
\end{itemize}

\section{LLM Characterization}
\label{sec:characterization}

In this section, we introduce modern LLMs and extensively characterize their power usage patterns at a server level, focusing on the intrinsic differences between training and inference workloads, prompt and token phases within inferences, and behaviors under GPU power management techniques.
\begin{figure*}
    \centering
    \subfloat[Flan-T5.]{
        \includegraphics[width=0.24\textwidth]{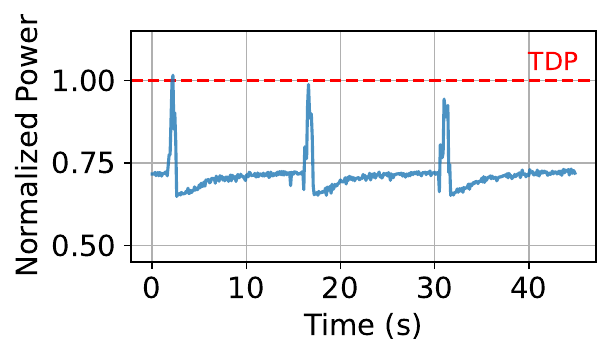}
        \label{fig:inference-timeseries-t5}
    }
    \subfloat[GPT-NeoX-20B.]{
        \includegraphics[width=0.24\textwidth]{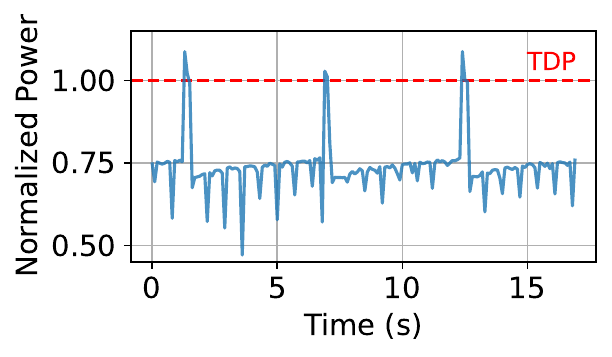}
        \label{fig:inference-timeseries-gpt}
    }
    \subfloat[OPT-30B.]{
        \includegraphics[width=0.24\textwidth]{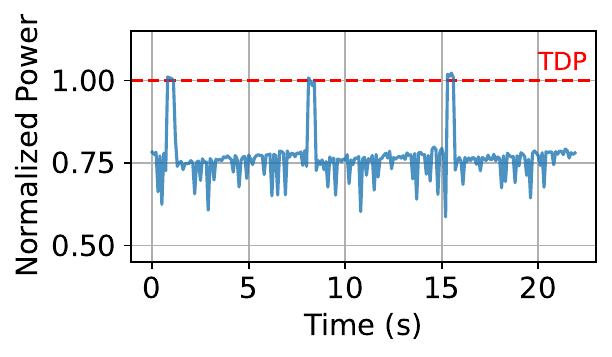}
        \label{fig:inference-timeseries-opt}
    }
    \subfloat[BLOOM-176B.]{
        \includegraphics[width=0.24\textwidth]{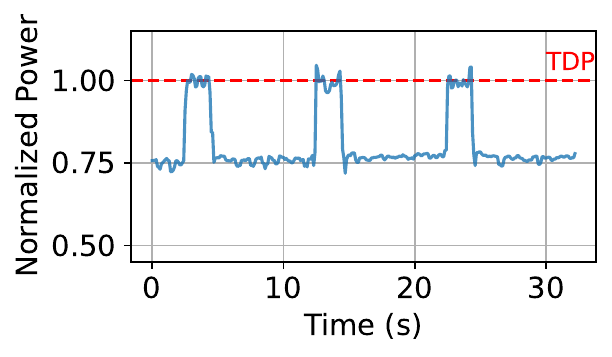}
        \label{fig:inference-timeseries-bloom}
    }
    \caption{
    GPU power usage timeseries for multiple inference models.
    It shows distinct power usage patterns in prompt (spiky) vs. token phase (longer, more stable, and lower).
    The phases in each model take different amount of times.
    }
  
    \label{fig:inference-timeseries}
\end{figure*}

\subsection{LLMs in the Cloud}

\myparagraph{Transformer models}
We focus this work on modern LLMs that are transformer-based.
Beyond the tokenizer and embedding layers, transformer models generally consist of attention~\cite{vaswani2017attention} and multi-layer-perceptron layers to contextualize the inputs, and generate an output.
\emph{Encoder-only} transformer models like BERT~\cite{devlin2018bert} and RoBERTa~\cite{liu2019roberta} consist of bi-directional self-attention, allowing them to contextualize the tokens altogether for language understanding tasks like summarizing and sentiment analysis.
\emph{Decoder-only} transformer models like GPT~\cite{radford2019gpt} and BLOOM~\cite{scao2022bloom} consist of masked or uni-directional self-attention for generating language sequences.
\emph{Encoder-decoder} models like FLAN-T5~\cite{chung202flant5} use an encoder for understanding the input and the decoder for generating text.
Transformer models can be for language, vision, or multi-modal.
Even though we focus on transformer-based LLMs, in \Cref{sec:discussion} we show that our work can be extended to vision and multi-modal models.%

\myparagraph{Training vs. inference}
LLMs are generally used in a train once and use forever mode. Training is generally much more resource intensive than inference, as the model is fed a lot of data for many iterations in parallel. For example, OpenAI scaled up their clusters to 7,500 GPU servers to train LLMs like GPT-3~\autocite{openai_7500k8s}.
Since training for LLMs is generally run across thousands of GPUs, and it must update weights across GPUs, it involves both a computation-heavy and a communication-heavy phase per iteration. Especially between the forward and backward passes in a training iteration, there is a communication bubble.
On the other hand, inference only requires forward pass of the model, and requires fewer resources. For instance, a BLOOM inference (similar in size to GPT-3) can be served using 8 GPUs on a single server.

\myparagraph{Prompt processing vs. token sampling}
\Cref{fig:prompt_v_token} shows the two main phases in an LLM inference: prompt processing, and token sampling. Prompt sampling can be done in parallel on all the input tokens, making it compute-intensive. On the other hand, token sampling is sequential, and uses the cached data from the tokens processed so far, making it a computationally light, and memory bandwidth-intensive phase.
Most of the computation, such as KV-cache, during the prompt processing phase is then cached to avoid recomputation~\autocite{vaswani2017attention}.

\myparagraph{Configuration knobs in training and inference}
\begin{itemize}
\item \emph{Batch size} defines the number of sequences processed together. A higher batch size leads to better throughput.
For training, the norm is to maximize the batch size based on the available memory and compute.
For inference, batching is best-effort, since waiting to fill the batch size can cause delays in the results.
\item \emph{Input size} defines the length of the prompt sequence.
\item \emph{Output size} defines the maximum number of output tokens to be generated per prompt. The model can generate end of sequence (EOS) before the suggested reaching the output size, thereby truncating the output. 
\end{itemize}

\myparagraph{Offering LLMs in the cloud}
Cloud providers can host LLMs in different way. First, the customers can bring their own models and host them using the infrastructure-as-a-service based virtual machine offerings, making the model opaque to the cloud provider. Second, the cloud provider could use platforms like Singularity~\cite{shukla2022singularity}, Azure OpenAI service~\cite{azure_openai}, Google BARD~\cite{google_bard}, or Azure ML~\cite{azureml} to help customers use known models. This method provides the cloud provider visibility into the model type.

\subsection{Characterization Methodology}
\label{sec:characterization_methodology}

\myparagraph{Hardware}
We run workloads on two NVIDIA DGX A100 machines with $8\times$A100-40GB and $8\times$A100-80GB GPUs respectively~\autocite{nvidia_dgx_a100,nvidia_a100}.
Due to GPU availability crunch, we use the former machine to run training workloads and the latter to run inference workloads.
GPUs communicate with the host CPU using PCIe 4.0 and are interconnected via NVLink 3.0 for fast inter-GPU communication. 
The CPU is a dual-socket AMD Rome.
\Cref{fig:A100Power} shows that GPUs make around 50\% of the server power.

\myparagraph{Workloads and metrics}
\Cref{tab:workloads} shows the LLMs we evaluated which span domains and tasks.
We consider popular open-source LLM models varying structures and sizes: Encoder (RoBERTa~\cite{liu2019roberta}), Decoder (GPT-NeoX, OPT, BLOOM~\cite{scao2022bloom}), and Encoder+Decoder (Flan-T5~\cite{chung202flant5}) transformer models~\autocite{vaswani2017attention}.
We use the models weights and training scripts from open-source repositories (\ie,  Huggingface Transformers~\autocite{wolf2020transformers}, GPT-NeoX library~\autocite{gpt-neox-library}, and DeepSpeed~\autocite{aminabadi2022deepspeed, deepspeed_mii, flan-t5_finetune}).

To accurately reflect inference efficiency, we use parallelization strategies including tensor and data parallelism, which are supported by popular open-source deep learning frameworks today~\autocite{deepspeed_mii,hf_accelerate,_PyTorch}.
To emulate the worst-case scenario in terms of power utilization, we saturate GPUs by running a constant stream of inference requests with no idle time.

For training, we run each workload on dedicated server for at least 5 minutes (100+ iterations) across all 8 GPUs. 
Our training setup uses distributed data parallelism (DDP) and/or tensor parallelism~\cite{li2020pytorch, shoeybi2019megatron}.
We configure the training batch sizes to use at least 85\% of the GPU memory.

We run NVIDIA DCGM with a 100 ms interval to track utilization, power consumption, temperature, hardware activity, and other performance counters on each GPU~\autocite{nvidia_dcgm}.

\myparagraph{GPU controls}
For our characterization, we use nvidia-smi based GPU controls on the A100 GPUs. The two main controls we explore are power caps and frequency caps. Each GPU supports power caps ranging from 100W–400W and SM clock frequencies from 0.2GHz–1.4GHz. For power capping and frequency scaling experiments, we evaluate a subset of the supported GPU clock frequencies and power caps ranging between 1.1–1.4GHz and 325–400W respectively to make the parameter space tractable.
We discuss a subset of results.

\subsection{LLM Inference Power Characterization}
\label{sec:inference}

\myparagraph{Phases in power consumption}
\Cref{fig:inference-timeseries} shows the time-series power consumption for multiple inference models, each with three inferences of the same prompt. We observe that across all inference models, during every iteration, the power usage patterns exhibit two distinct phases: power spikes in the beginning, and a stable, lower power consumption later.
Power spikes consistently occur at the start of every inference request, often going beyond GPU's TDP values. These spikes correspond to the compute-intensive prompt phases of LLMs as all the token in the prompt can be processed in parallel, causing a large input matrix (\ie, the input query to the model).
Following the spike, the stable, lower power consumption phase corresponds to sequential, autoregressive token sampling.
The token sampling phase sequentially generates new tokens by re-using activations stored in the KV-cache and only incurs light computation, so the power draw during token sampling phase relatively stable and low.
Since a large number of output tokens may be sequentially generated in response to a single request (an input prompt), the prompt-phase computation tends to last much shorter than the token phase and the resulting power spike per request generally lasts < 1 second.
Although we see these power consumption spikes across the GPUs serving the same inference, the spikes are not correlated across different endpoints serving different inferences. This is because of the arrival time variation in the prompts and scheduling. Therefore, at a cluster level, we would not expect to see these peaks align. Instead, a statistical multiplexing of the prompt and token processing stages across various servers should yield lower peak power consumption at the cluster level, despite high peak power numbers per server. 

\begin{figure}
    \centering
    \subfloat[Power varying input sizes.]{
        \includegraphics[width=0.48\columnwidth]{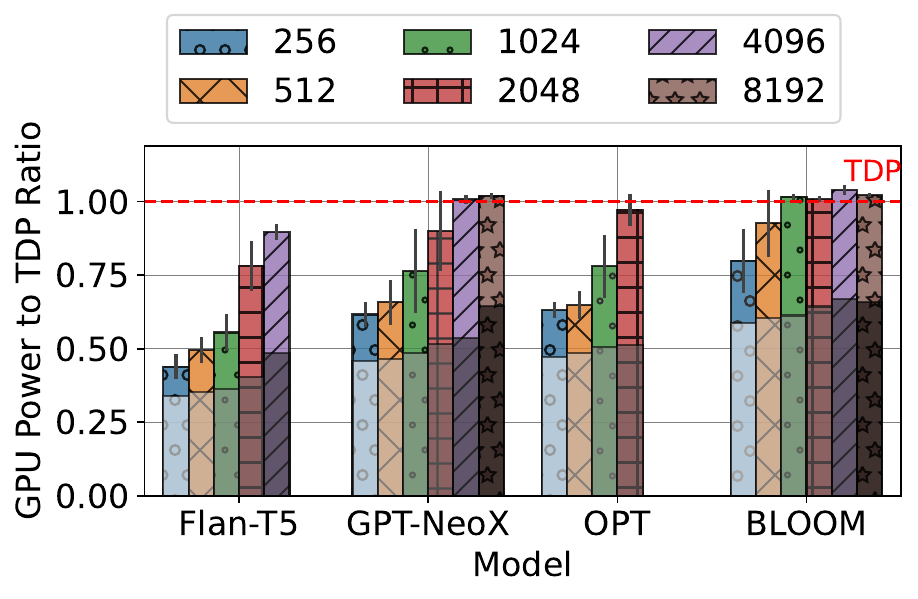}
        \label{fig:sensitivity_input_sizes_power}
    }
    \subfloat[Latency varying input sizes.]{
        \includegraphics[width=0.48\columnwidth]{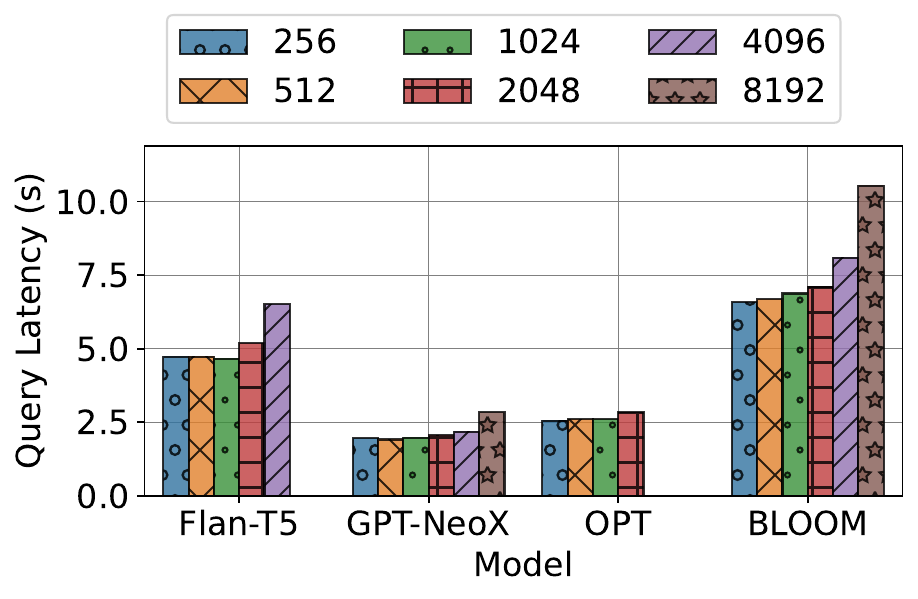}
        \label{fig:sensitivity_input_sizes_latency}
    }
    \\
    \subfloat[Power varying batch sizes.]{
        \includegraphics[width=0.48\columnwidth]{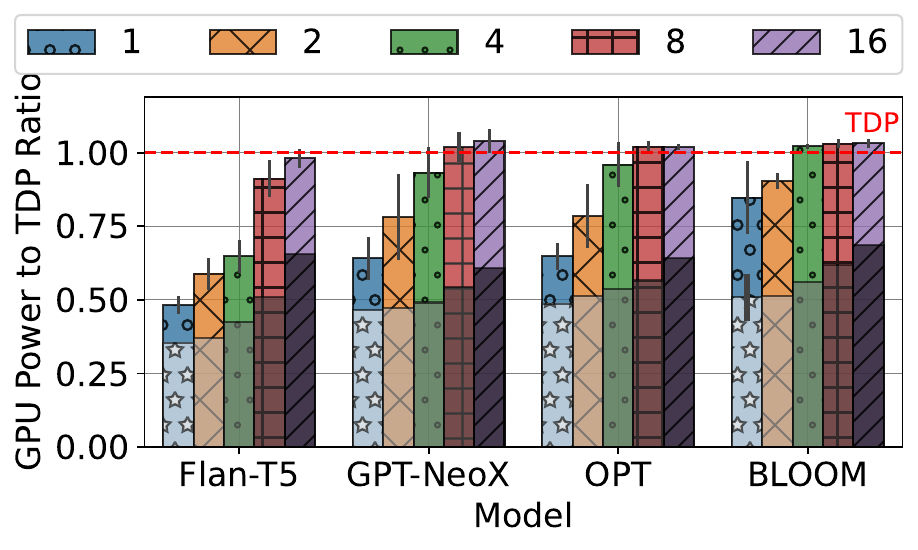}
        \label{fig:sensitivity_batch_sizes_power}
    }
    \subfloat[Latency varying batch sizes.]{
        \includegraphics[width=0.48\columnwidth]{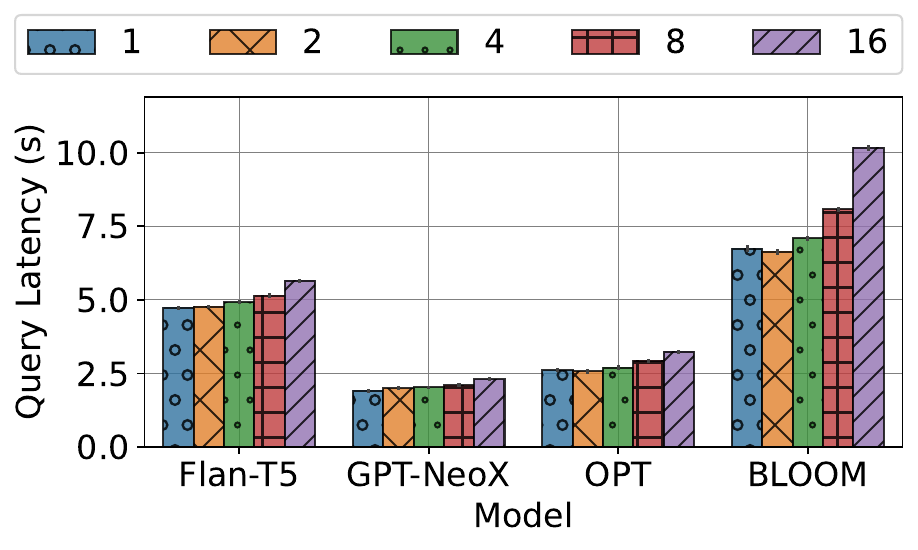}
        \label{fig:sensitivity_batch_sizes_latency}
    }
    \\
    \subfloat[Power varying output sizes.]{
        \includegraphics[width=0.48\columnwidth]{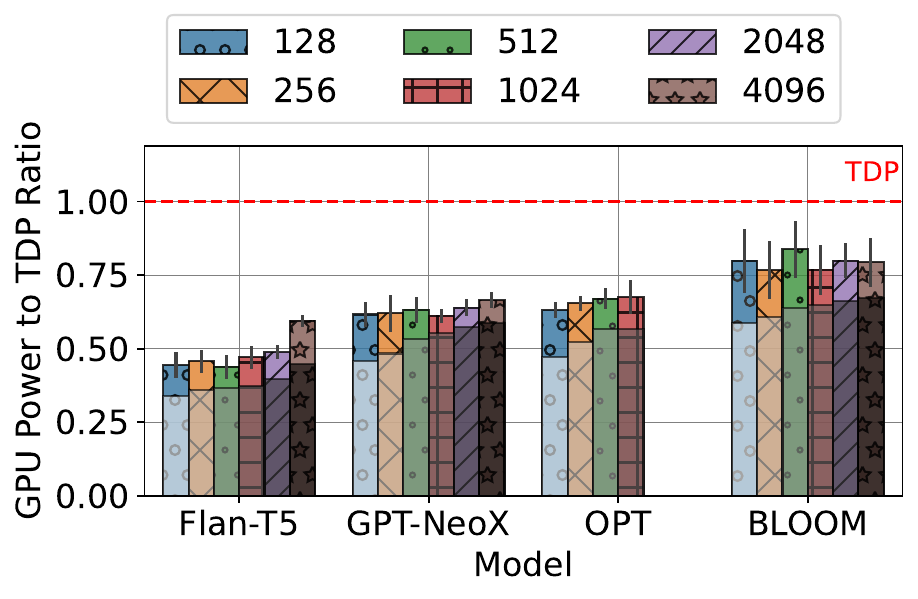}
        \label{fig:sensitivity_output_sizes_power}
    }
    \subfloat[Latency varying output sizes.]{
        \includegraphics[width=0.48\columnwidth]{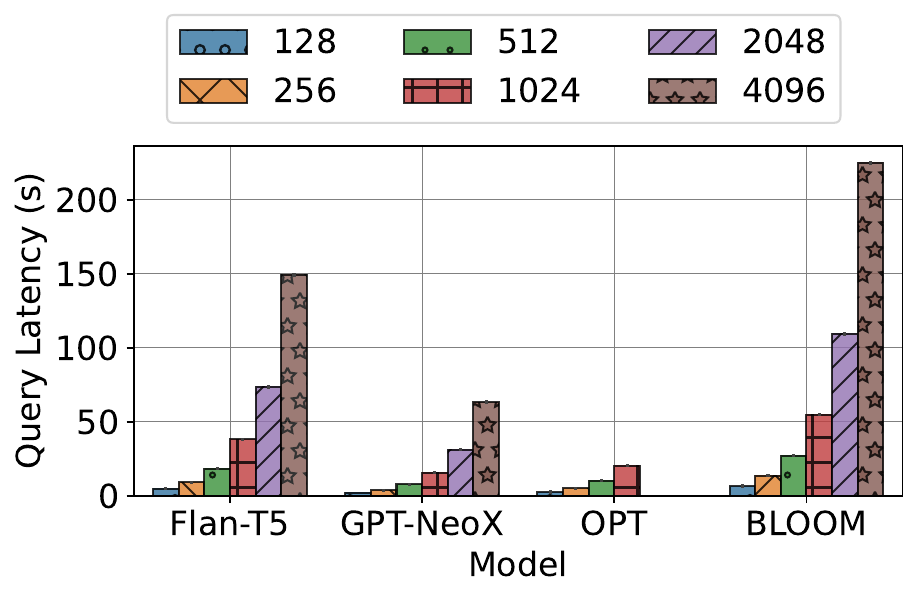}
        \label{fig:sensitivity_output_sizes_latency}
    }
    \caption{Power (mean, peak) and latency sensitivity to the input, batch, and output sizes for multiple inference models running on A100-80GB GPUs.
    }
    \label{fig:sensitivity_inference} 
\end{figure}

\myparagraph{Power patterns with different configurations}
Inference workloads can differ in their input and output parameters based on the use cases, effectively changing the amount of computation during prompt and token phases. Therefore, we closely examine the power consumption patterns and latency implications for a variety of LLM inference configurations in~\Cref{fig:sensitivity_inference}. To separately characterize the peak and mean power, we depict GPU power normalized to TDP for each model and configuration in stacked bars of two components: the lower (opaque bars --- mean power during iterations) and the higher (regular bars --- peak power). 
We first observe that with the same configurations, larger models (\eg, BLOOM-176B) incur a much higher amount of computation during both prompt and token phases, and show significantly larger peak and mean power consumption.

\myparagraphemph{Input sizes}
\Cref{fig:sensitivity_input_sizes_power} shows the mean and peak power to TDP ratio varying input sizes from 256 to 8192 tokens.
As input size increases, peak power drastically goes up across all models, reflecting the significant increase in prompt phase computations. The mean power, dominated by lighter token sampling computation, remains stable and low, further drawing the distinction of power patterns between prompt and token phases. \Cref{fig:sensitivity_input_sizes_latency} shows the corresponding latency results. As the sequential token sampling phase contributes to most of the query latency, increasing input sizes shows little effect on latency until >4k input tokens.

\myparagraphemph{Batch Sizes}
\Cref{fig:sensitivity_batch_sizes_power} shows the power impact of varying batch sizes from 1 to 16.
Larger batch sizes effectively increase the input sizes for prompt computation, resulting in similar increase to peak power draws. Mean power also exhibits a gradual increase, since the effective number of tokens processed concurrently during token phase is higher.
\Cref{fig:sensitivity_batch_sizes_latency} shows a slight increase in latency due to computation increases in both prompt and token phases from larger batch sizes.

\myparagraphemph{Output Sizes}
Because the sequential and autoregressive nature of token sampling, similar computation and power consumption patterns are expected to repeat for each generated token. As a result, \Cref{fig:sensitivity_output_sizes_power} shows that increasing output size does not affect the peak and mean power drawn but simply increases the duration of request execution linearly (\Cref{fig:sensitivity_output_sizes_latency}).

\begin{figure}
    \centering
    \subfloat[No cap.]{
        \includegraphics[width=0.32\columnwidth]{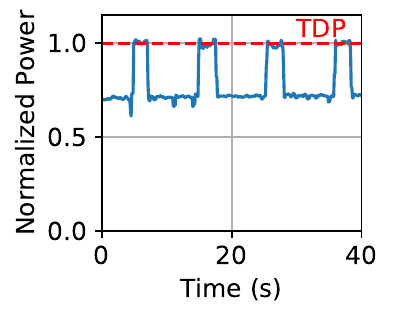}
        \label{fig:inference_nocap}
    }
    \subfloat[325W cap per GPU.]{
        \includegraphics[width=0.32\columnwidth]{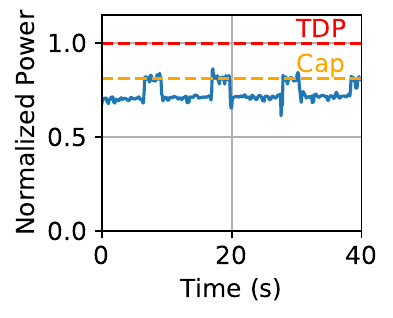}
        \label{fig:inference_powercap}
    }
    \subfloat[1.1GHz SM clock.]{
        \includegraphics[width=0.32\columnwidth]{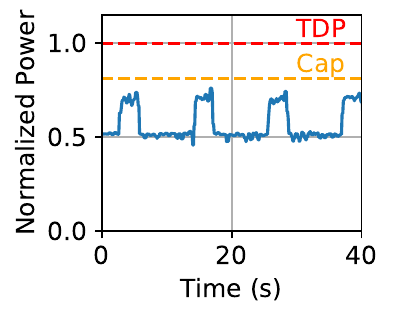}
        \label{fig:inference_freqcap}
    }
    \caption{GPU power capping and frequency scaling on BLOOM inference (input=8192, output=128, and batch=1).}
    \label{fig:inference_nocap_vs_powercap_vs_freqcap}
\end{figure}

\begin{figure}
    \centering
    \subfloat[All Models.]{
        \includegraphics[width=0.49\columnwidth]{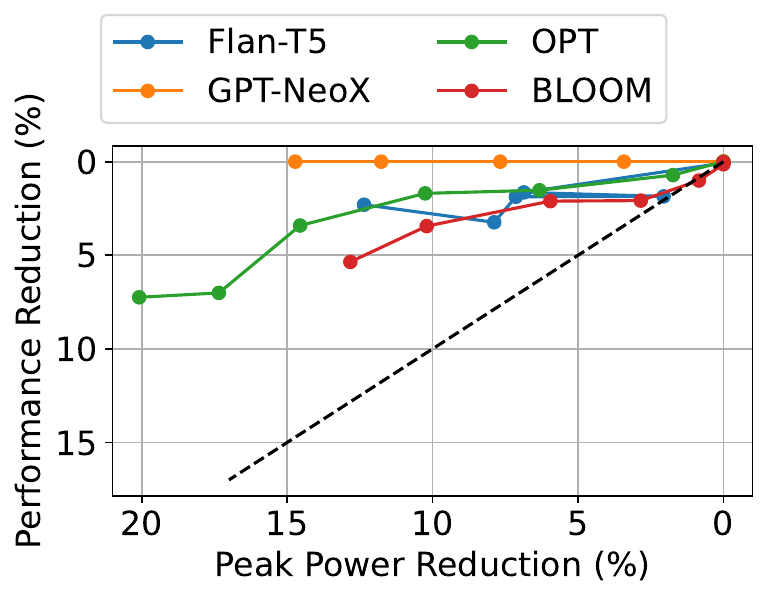}
        \label{fig:inference_llm_all}
    }
    \subfloat[BLOOM sensitivity.]{
        \includegraphics[width=0.49\columnwidth]{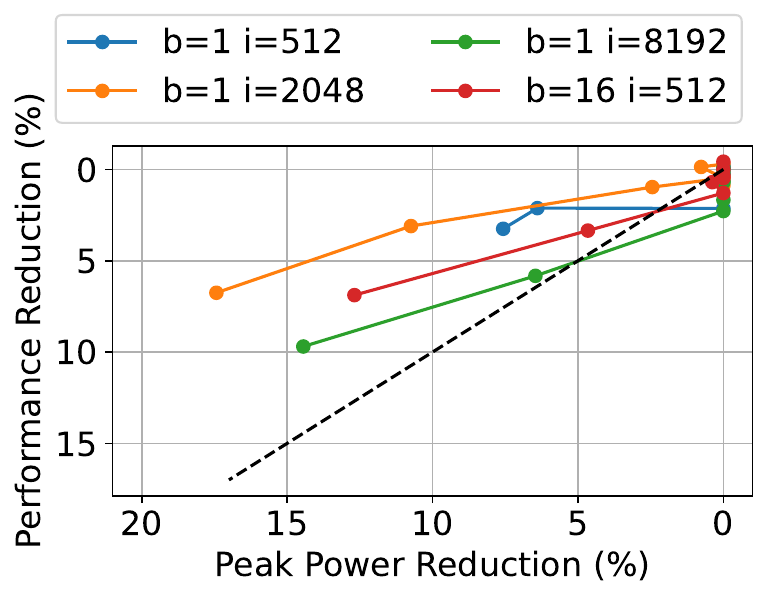}
        \label{fig:inference_llm_bloom}
    }

    \caption{Peak power reduction (based on TDP) vs. performance reduction for multiple inference models at varying GPU SM frequencies. The dashed black line shows a linear scaling of performance with power drawn.
    }
\label{fig:inference_peak_power_vs_perf}
\end{figure}

\myparagraph{Impact of frequency and power capping}
\label{sec:perfcharacterizeInference}
As discussed in \cref{sec:characterization_methodology}, the two main controls we can use in GPUs today are frequency and power capping. 
\Cref{fig:inference_nocap_vs_powercap_vs_freqcap} shows the impact of capping the power vs frequency on BLOOM inference. Since power capping is reactive, it allows the initial peaks in the prompt phase to go beyond the power cap. Frequency cap being proactive is a better control, but leads to performance impact throughout the execution, and not just when the power utilization is high. Keeping this in mind, we should choose frequency capping for a more reliable control. We discuss this further in \cref{sec:enabling_oversubsciption}.

The next step is to quantify the performance and peak power impact of frequency capping. \Cref{fig:inference_llm_all} shows the relative peak power and performance reduction compared to no capping by varying GPU clock frequencies for all models.
We observe that the relationship between power reduction and performance is superlinear---significant power (up to 20\%) can be reclaimed for minimal performance loss (maximum up to 7\%).
Notably, the sensitivity to peak power reduction mechanisms varies across different models. In particular, larger models tend to have more performance impact from frequency capping. For example, GPT-NeoX incurs no performance loss while BLOOM exhibits 5\% at a similar peak power reduction level (13\%). \Cref{fig:inference_llm_bloom} further shows the sensitivity results of varying prompt computation (input and batch size) for BLOOM. Smaller total input size shows less performance loss with the same amount of peak power reduction, because there is less prompt phase computation that gets impacted by reduced frequencies.
Overall, across models and configurations, the peak power reduction from locked frequency execution is substantially higher than the relative performance drop.

\subsection{LLM Training Power Characterization}
\label{sec:training}

\begin{figure}[t!]
    \centering
    \includegraphics[width=\columnwidth]{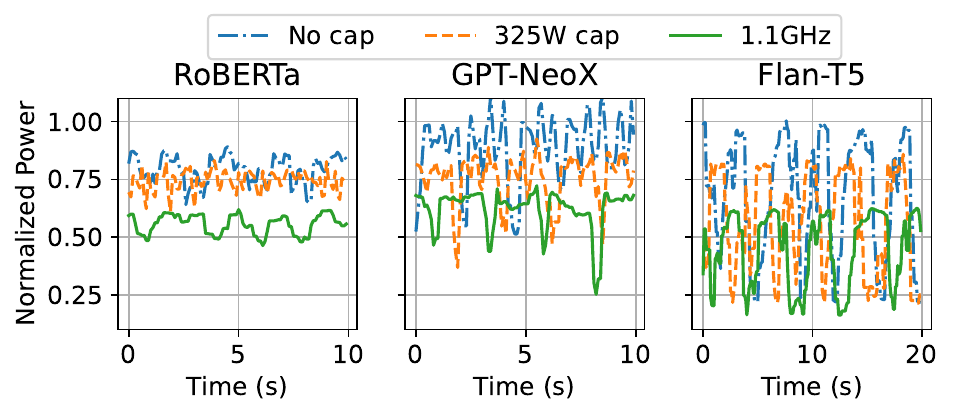}
    \caption{Power usage timeseries for training workloads under no cap, power cap, and frequency cap. 
    }
    \label{fig:training_power_swings_with_freq_scaling}
\end{figure}

\begin{figure}[t!]
    \centering
    \subfloat[SM freqs.]{
        \includegraphics[width=0.5\columnwidth]{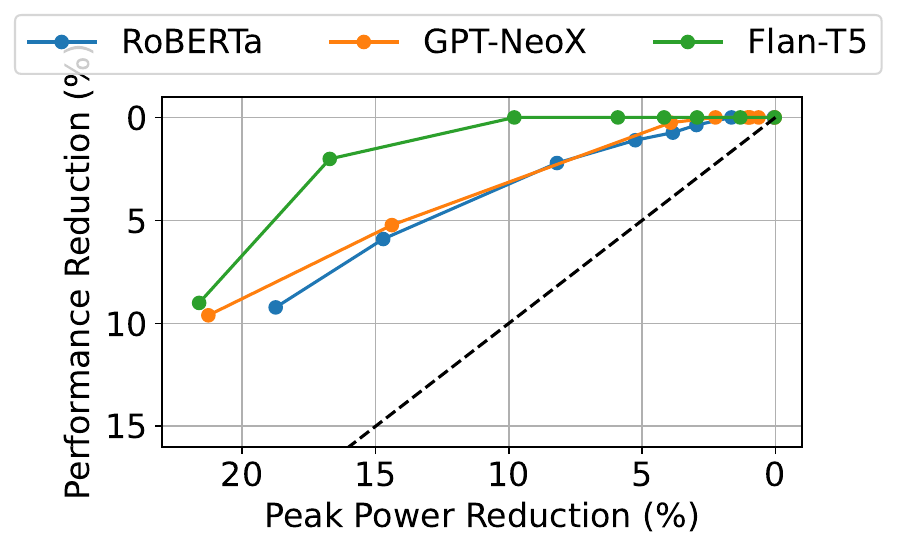}
    }
    \subfloat[Power caps.]{
        \includegraphics[width=0.5\columnwidth]{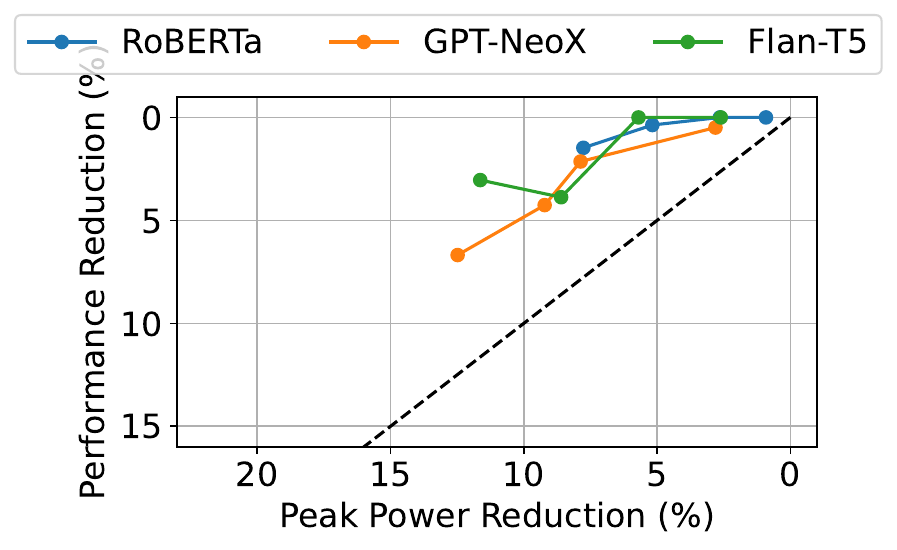}
    }
    \caption{Peak power vs. performance reduction for training.
    }
    \label{fig:training_peak_power_vs_perf}
\end{figure}

\myparagraph{Peak power}
\Cref{fig:training_power_swings_with_freq_scaling} (blue) shows the time series peak power data updated every 100ms for 5 iterations of training per model. 
The peak power during the training iterations goes up to the TDP of the GPUs, and beyond for GPT-NeoX and Flan-T5. On the other hand, RoBERTa, an encoder-only model does not reach the TDP. Note that different types of data sharding, batching, and parallelism techniques could slightly change this behavior. Our takeaway here is that training can easily reach the TDP of the system.

\myparagraph{Power swings}
Interestingly, \Cref{fig:training_power_swings_with_freq_scaling} (blue) also shows that there are big swings in power draw across GPUs each iteration. For example, in RoBERTa, an iteration lasts for ~1 second. Each iteration has a small dip in power around the 0.5 second mark, and a big dip in power at the end. Based on the model architecture, it is clear that the smaller dip is caused between the forward and backward propagation phases, as the threads working on the same data synchronize, and the GPU utilization decreases. The larger dip on the other hand, is caused at the end of the iteration as all the GPUs synchronize before the next iteration starts. Therefore, the power swings are caused by the inherent workload behavior of switching between computation- and communication-intensive phases.

The power consumption in the communication-intensive phase is different across the three models. While RoBERTa is still at 75\% of the TDP at the iteration boundary, GPT-NeoX drops down to 50\%, and Flan-T5 goes down all the way to 20\%, which corresponds to the idle power of the GPUs.

In a larger scale training, these power swings will be correlated across thousands of GPUs working on the same training job, potentially causing challenges in the power delivery infrastructure. 
Note that the main challenge here is the power swings, and not the peak power.
\begin{table*}[btp]
\begin{minipage}[b]{0.27\textwidth}
    \centering
    \includegraphics[width=1\textwidth]{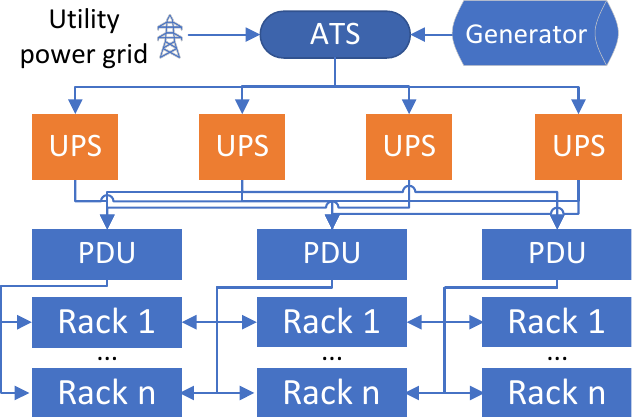}
    \captionof{figure}{Example of the power architecture in a datacenter.}
    \label{fig:dc_power}
\end{minipage}
\hfill
\begin{minipage}[b]{0.30\textwidth}
    \centering
    \includegraphics[width=0.9\textwidth]{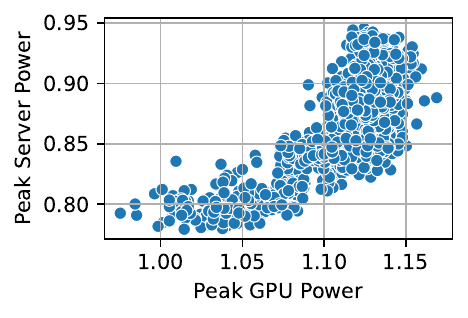}
    \captionof{figure}{Server and GPU peak power consumption normalized to their TDPs.}
    \label{fig:server_gpu_scatter}
\end{minipage}
\hfill
\begin{minipage}[b]{0.30\textwidth}
    \centering
    \footnotesize
    \begin{tabular}{cc}
    \toprule
    \textbf{Parameter}                  & \textbf{Value} \\
    \midrule
    \textbf{Number of servers}          & 40   \\
    \textbf{Server type}                & DGX-A100 \\
    \textbf{Power telemetry delay}      & 2s \\
    \textbf{Power brake latency}        & 5s  \\
    \textbf{OOB commands latency}       & 40s \\                                       
    \bottomrule
    \end{tabular}
    \captionsetup{skip=25pt}
    \caption{Default row-level parameters for our study.}
    \label{tab:default_cluster_params}
\end{minipage}
\end{table*}
\label{sec:clusterpower}

\myparagraph{Impact of capping}
We investigate the impact of frequency capping on the training workloads and their power consumption patterns.
\Cref{fig:training_power_swings_with_freq_scaling} shows the power consumption under power and frequency caps. The peak power is reduced by up to 20\% for these workloads. However, our main challenge is the power swing, for which, we will need to bring down the peak power, while maintaining the power troughs high. Note that RoBERTa and GPT-NeoX having a considerable power consumption at the iteration boundary means that they have computations in the GPUs even in that phase. Therefore, for these models, capping brings down the trough in the power consumption too, not helping the power swing challenge. On the other hand, Flan-T5 training iteration boundary brings the GPUs down to idle, thereby reacting well to capping. 

\Cref{fig:training_peak_power_vs_perf} shows the impact of frequency and power capping on the throughput of the training. Frequency capping is more effective in reclaiming larger amount of peak power, and power capping introduces more performance variability with less control over computation. For Flan-T5 and GPT-NeoX, frequency capping reduces the peak server power by 22\% while only impacting the performance by 10\%. For single large cluster-level training jobs, this can be used to deal with the power swings, or, in case of multiple smaller concurrent jobs, can be used to oversubscribe power for more deployments.

Since the usage of a training cluster over its lifetime can vary in terms of the training job sizes, power oversubscription could mean turning the cluster partially off to support larger training jobs. Given the high cost and low availability of GPUs, we think this is not a reasonable approach. Therefore, we conclude that power oversubscription in training clusters is tricky. Instead, we think that our capping results can mainly be used to deal with the power swings effectively.

\section{LLM Cluster Power Characteristics}

\begin{table}[t!]
\centering
\footnotesize
\begin{tabular}{ccc}
\toprule
 & \textbf{Training} & \textbf{Inference}   \\ \midrule
\textbf{Peak power utilization}  & 97\%  & 79\%               \\
\textbf{Power usage pattern}                                                                     & \begin{tabular}[c]{@{}c@{}}Coordinated swings \\ every few seconds\end{tabular} & \begin{tabular}[c]{@{}c@{}}Diurnal with \\ short-term variations\end{tabular} \\
\textbf{Max. power spike in 2s}                                                                  & 37.5\%                                                                          & 9\%                                                                           \\
\textbf{\begin{tabular}[c]{@{}c@{}}Max. power spike in 5s\\ (power brake latency)\end{tabular}}  & --                                                                              & 9.1\%                                                                         \\
\textbf{\begin{tabular}[c]{@{}c@{}}Max. power spike in 40s\\ (OOB capping latency)\end{tabular}} & --                                                                              & 11.8\%                                                                        \\
\bottomrule
\end{tabular}
\caption{LLM cluster power usage in production.}
\label{tab:cluster_power_usage}
\end{table}

\subsection{Datacenter Power Management}

\myparagraph{Power provisioning}
GPU servers by default are provisioned for peak power draw because:
(1)~GPUs are typically designed to maximize FLOPS, so hitting peak power draw is a likely scenario,
(2)~cloud servers may run any workload, so provisioning for the worst case ensures safety, and
(3)~out-of-band GPU power monitoring is slow and capping not very reliable, which makes power oversubscription difficult~\autocite{patel2023towards}.
Consequently, provisioning power for GPU servers is expensive.

Most large-scale CPU clusters deployed today use some form of power oversubscription to reduce cost~\autocite{kumbhare2021prediction,fu2011much}.
For example, they might uniformly de-rate servers, use workload-aware power capping, or implement throttle-aware power management.
In contrast, power oversubscription is challenging in GPU clusters, as we elaborate upon in \Cref{sec:challenges}.

A datacenter floor plan is generally built around the power hierarchy.
A few servers are deployed within a rack, and several racks make a row in the datacenter.
\Cref{fig:dc_power} shows an example power hierarchy, where the PDUs would power the row of racks~\cite{zhang_Flex}.

\subsection{GPU Power Usage Patterns at Scale}

We only show subsets of the data from production clusters and normalize the numbers for confidentiality.%

\myparagraph{Row-level Power}
\Cref{tab:cluster_power_usage} shows the normalized aggregate power consumption patterns of LLM training and inference clusters at a large cloud provider.
Note that we consider an interactive inference cluster (\ie, where users make inference requests and expect rapid responses).
We observe that:
(1)~training has higher peak and average power draw compared to inference,
(2)~training incurs large swings in power consumption, up to 37.5\% of the provisioned power capacity, whereas inference only incurs a change of up to 9\%,
(3)~inference power consumption shows a diurnal pattern since it is an interactive workload; yet, over the course of a few seconds, its power usage remains relatively stable compared to training.

These differences imply that training tends to put much higher strain on the cluster power delivery infrastructure compared to inference. The power swings at scale are a challenge, as we also predict in \cref{sec:characterization}. Inference workloads are promising for power oversubscription, whereas training workloads are not.

\myparagraph{Server-level power}
Next, since peak power drives power provisioning decisions per server, we plot the peak server and GPU power in a production cluster, relative to their TDP, in~\Cref{fig:server_gpu_scatter}.
We find that:
(1) GPU power constitutes on average 60\% of server-level power consumption; hence we focus on GPUs in the rest of this paper,
(2) peak GPU power far exceeds the overall server GPU TDP (by up to 500W),
(3) peak server power is highly correlated with peak GPU power,
(4) peak GPU power has a much smaller range than peak server power, and
(5) peak power remains largely unchanged over time since servers are heavily utilized.

\section{Challenges}
\label{sec:challenges}
Power oversubscription in LLM clusters today poses several challenges.

\myparagraph{A. Mixed inference and training}
Current datacenters host both inference and training in the same infrastructure.
However, when managing power this is suboptimal due to the huge disparity between their at-scale characteristics as discussed in \cref{sec:clusterpower}.
Power swings in large training jobs preclude power optimizations for inference.

\myparagraph{B. Latency sensitive workloads}
Servers in the LLM clusters could be deployed for various use-cases.
While use-cases like summarization and understanding tasks may not be latency-sensitive, a use-case like chat is very latency-sensitive.
Capping these all workloads equally is unreasonable and we should prioritize latency-sensitive ones.

\myparagraph{C. Distinct inference phases}
Based on the power characterization in \cref{sec:characterization}, LLM inference has two distinct phases with very different profiles.
While the prompt processing takes up a lot of power, token sampling does not.
This makes it really difficult to manage power for these workloads.

\myparagraph{D. Virtualized environment}
Cloud providers can host LLMs in Virtual Machines (VMs) or as services like Amazon SageMaker~\cite{awssagemaker}, Azure ML~\cite{azureml}, or Google Vertex AI~\cite{googlevertexai}.
In all of these cases, GPUs are used in Direct Device Access (DDA) mode, that precludes the cloud provider from accessing the GPU drivers.
Reliable and fast power or frequency capping is necessary as a fallback for power oversubscription, in case the overall power draw exceeds supported capacity.
Although GPUs have fast in-band controls (for instance, nvidia-smi for A100 GPUs) that allow the use of drivers for frequency and power capping, these are out of reach for the cloud provider.

\myparagraph{E. Tight latency bounds with slow interfaces}
At a worst-case load of 133\%, UPSes provide 10\s for failure tolerance~\autocite{zhang_Flex}.
This places an upper bound on the latency of any backstop capping, and makes it a requirement that the control happens out-of-band, without the VM in the loop.

\myparagraphemph{Lack of server-power controls}
Features like Intel RAPL~\cite{pavlos2015_powercapping} provide fast and reliable controls in the CPU servers to bring down the power of the entire server down, by setting a single cap on the CPU. However, for GPU servers, where about 60\% of the consumed power comes from GPUs, there are no fast controls available to bring the entire server power down by setting one cap.

\myparagraphemph{Lack of fast out-of-band controls}
Due to challenge D, the virtualization and DDA, any power controls need to be accessible out-of-band, through the rack manager and board management controller (BMC) to be useful for cloud providers. 

NVIDIA has a few controls through the SMBPBI~\cite{nvidia_smbpbi} that are helpful, but slow. The available controls are: (1) Frequency caps for the GPU compute. Note that this does not allow us to control the GPU memory clock, but just the GPU compute clock. (2) Power caps at the GPU level, which allow us to cap the power consumed by individual GPUs. As shown in \cref{sec:characterization}, the power caps do not guarantee the spikes from breaching the desired level. (3) Power brake, which is a fast lever (\cref{tab:default_cluster_params}) to bring the GPU down to almost a halt, stopping all progress.

\myparagraph{F. Changes in the workload over time}
The number of additional racks added through power oversubscription is a static decision, that stays for the lifetime of the servers (\ie, 4-6 years). However, as the types of models and their use-cases running in the cluster change with time, the policy for power capping needs to be configurable enough to support high performance for the workloads at most times, and avoid frequent power capping.

\myparagraph{G. Reliability}
The power and frequency capping are backstops to ensure that the UPS does not trip. 
Therefore, the path exercised to implement these need to be extremely reliable.
Therefore, we need to avoid depending on the user or any unreliable code during the critical path for capping.

\myparagraph{H. Ability to work with existing datacenters}
Any changes to the datacenter architecture would take many years to be available.
Such solutions are impractical to support the current LLM demand.
Our solution needs to work with existing infrastructure available today.
For this, we need to be able to retrofit into the existing datacenters.

\begin{table*}[ht]
\begin{minipage}[b]{0.33\textwidth}
\centering
\scriptsize
\begin{tabular}{@{}ccc@{}}
\toprule
\textbf{Mode} & \textbf{Low Priority} &\textbf{High Priority}                             \\ \midrule
\textbf{Uncapped}  & Uncapped  & Uncapped     \\
\textbf{Threshold T1} & Frequency capped  & Uncapped\\
&(1275 MHz)&\\
\textbf{Threshold T2}  & Frequency capped & Frequency capped\\
&(1110 MHz) & (1305 MHz) \\
\textbf{Powerbrake} & Frequency capped& Frequency capped\\ 
&(288MHz)&(288MHz)\\
\bottomrule
\end{tabular}
\caption{Power modes for low and high priority workloads.}
\label{tab:thresholds}
\end{minipage}
\hspace{0.5em}
\begin{minipage}[b]{0.23\textwidth}
    \centering
    \includegraphics[width=0.5\textwidth]{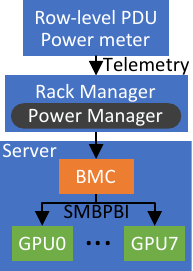}
    \captionof{figure}{Power management flow.}
    \label{fig:cluster_power_manager}
\end{minipage}
\hspace{0.5em}
\begin{minipage}[b]{0.35\textwidth}
\centering
\scriptsize
\begin{tabular}{@{}ccccc@{}}
\toprule
\textbf{Workload} & \textbf{Prompt size} & \textbf{Output size} & \textbf{Ratio} & \textbf{Priority} \\ \midrule
Summarize         & 2048-8192            & 256--512             & 25\%           & Low               \\
Search            & 512--2048            & 1024--2048           & 25\%           & High              \\
Chat              & 2048--4096           & 128--2048            & 50\%           & 50:50             \\ \bottomrule
\vspace{1.5em}
\end{tabular}
\caption{Workloads distribution, based on the BLOOM-176B model.}
\label{tab:workload_distribution}
\end{minipage}
\end{table*}

\section{Designing \papername}
\label{sec:enabling_oversubsciption}
We start by discussing system-level design points and decisions addressing aforementioned challenges, followed by the detailed policy and flow.

\begin{table}
\centering
\scriptsize
\begin{tabular}{@{}ccc@{}}
\toprule
\textbf{Metric} & \textbf{High priority workload} & \textbf{Low priority workload} \\ \midrule
P50 latency impact         & < 1\%            & < 5\%  \\
P99 latency impact            &  < 5\%           &  < 50\% \\
Number of powerbrakes              & 0           & 0\\ \bottomrule
\end{tabular}
\caption{Service level objectives (SLOs) for \papername.}
\label{tab:SLO}
\end{table}

\myparagraph{A. Inference-optimized clusters}
Challenge A makes it difficult to oversubscribe power in clusters today due to the mix of inference and training.
We argue that there is a need to have separate clusters, optimized for inference, since the demand for inference is much higher (more than 90\%) than large training jobs~\autocite{patterson2022carbon,aws_inferentia_instances,tirias2019why}.
We build \papername targeting inference-optimized clusters, with room for further optimizations, like removing the back-end infiniband or other RDMA network in the inference clusters.

\myparagraph{B. Per-priority power capping}
To deal with challenge B, the latency-sesitive workloads, we propose two service-level priorities: high priority (HP) and low priority (LP), with the LP workloads being more probable to be capped. \cref{tab:workload_distribution} shows our example distribution of the priorities between the types of services.
The allocator in the cloud is aware of these workload priorities, and can make power-oversubsciption aware allocation to ensure a good mix of high and low-priority jobs in every row.

\myparagraph{C. Dealing with distinct inference phases}
Challenge C occurs due to the distinct power usage patterns between the prompt and token phases in LLM inference.
However, at a cluster-level, the statistical multiplexing of these phases reduces the power utilization peaks lower, as seen in \Cref{sec:clusterpower}.
With this insight, we choose a higher power aggregation level, the PDU breaker as our capping decision point.
This corresponds a row of racks as shown in \Cref{fig:dc_power}.

\myparagraph{D. Handling virtualization}
To provide guarantees against power trips, \papername uses only out-of-band interfaces that are available to the cloud providers from outside the VM.
We ensure that any of the settings we use can overwrite any settings that the VM user asks for, thereby dealing with challenge D.

\myparagraph{E. Designing within latency bounds}
\cref{tab:default_cluster_params} quantifies the latency of the slow out-of-band interfaces, representing challenge E.
The main latency upper bound is imposed by the 10\s deadline from the UPSes~\autocite{zhang_Flex}.
The PDU telemetry-based detection of a power threshold breach can be in the order of 3-5\s.
Given that the A100 powerbrake takes 5\s to implement, we meet the 10\s deadline from the UPSes~\autocite{zhang_Flex}.
However, powerbrake substantially throttles workload performance since it brings down the frequency of all the GPUs to 288 MHz, and should only be used in dire situations.
On the other hand, the less aggressive frequency and power caps take as long as 40\s to take effect. 
Our policy therefore uses multiple power thresholds, while accounting for any power spikes that may happen within these 40\s.
We track and use parameters like maximum, P99, and P90 power spikes at 40\s and 5\s granularity in our policy making.
\cref{tab:cluster_power_usage} shows a subset of these parameters.

\subsection{Policy: Thresholds and Power Modes}
\label{sec:POLCApolicy}
The goal of \papername is to maximize the additional servers deployed using power oversubscription, while meeting the Service Level Objectives (SLOs) in \cref{tab:SLO}.
To support per-priority performance SLOs, in \papername, we use two power thresholds, as shown in \Cref{tab:thresholds}.

\myparagraph{Threshold T1}
This is the lower power threshold, only applicable to low-priority workloads.
The two objectives here are: 
(1) to sufficiently avoid capping HP workloads, 
and (2) to do so while maintaining the SLOs for the low-priority workloads.
As we previously note in \Cref{sec:characterization}, prompt phase has high power peaks, while token phase does not.
A power cap only impacts the prompt phase, whereas a frequency cap reduces the power in both the phases. In order to maximize the power savings from capping low-priority workloads, we choose frequency capping for T1.
Upon reaching T1, we set all the low-priority workloads to the base frequency (the minimum promised frequency) of A100 GPUs, 1275 MHz.

\myparagraph{Threshold T2}
This is the upper power threshold, which is chosen to avoid powerbrakes completely.
We therefore use the observed value of maximum power spike in 40s to choose this threshold. 
When T2 is breached, we start by frequency capping all the low-priority workloads further down to 1110 MHz.
If the power is still above the threshold, we cap the HP workloads down to 1305MHz frequency, to incur negligible performance impact while still reclaiming power.

\myparagraph{Uncapping}
The policy also needs to define when to uncap the servers.
It is important to build in a hysteresis, to avoid constant capping, uncapping and overwhelm the power management system. Based on our parameter sweep, we choose the uncap thresholds to be 5\% below the corresponding capping threshold of T1 or T2. 

\myparagraph{Robustness and configurability}
We divide the challenge F (\ie, updates in LLMs over time) into two parts.
First, frequent efficiency and use-case updates to the models, that can cause their peak power to increase, while keeping the overall characteristics similar.
Second, long term changes in the model types.
For the first challenge, we build in robustness into \papername as we discuss and evaluate further in \Cref{sec:evaluation}.
For the second challenge, \papername infrequently updates the policy parameters (T1, T2, and the capping frequency) based on the workload changes over time.
For this, \papername tunes the variables using power traces combined with the history of capping decisions.
The steps followed for tuning are the same as we show in \Cref{sec:evaluation}.

\subsection{Power Management Flow}
We design a flow of control that is \textbf{reliable} and \textbf{works in existing datacenters}, therefore resolving challenges G and H.
\Cref{fig:cluster_power_manager} shows the hierarchy of telemetry and control in the power management in \papername.
The power manager running at rack-level receives frequent telemetry from the PDU about row-level telemetry.
We assume a homogeneous distribution of power and caps for fast control.
Based on its knowledge of the high vs low priority per VM, the power manager implements the threshold and caps as per the policy we describe in \cref{alg:cap}.
Once the BMC at the server gets the per-GPU caps from the power manager, it sets it across the GPUs in the server, using an interface like the SMBPBI~\cite{nvidia_smbpbi}.

The \papername design can be retrofitted into existing datacenters, without new hardware, meters, or structures.

\begin{algorithm}
\scriptsize
\caption{\papername power management.}
\label{alg:cap}
\begin{algorithmic}
\State $T1, T2 : $Power thresholds
\State $T1Buffer, T2Buffer : $Thresholds to uncap based on estimated power saves and spikes
\State $LP, HP : $Current workload priority fraction
\State $N :$ Number of servers in the row
\State $T1cap \gets false, T2cap \gets false, Powerbrake \gets false$
\Loop 
\State $P \gets Normalized Row Power Reading$
\If{$P > 1.0$}
    \State $Powerbrake \gets true$ \Comment{BMC sets powerbrake}
    \State $T1cap \gets true$
    \State $T2cap \gets true$
 \ElsIf{$P > T2$}
    \If{$T2cap == false$}
        \State $T2cap \gets true$
        \State LP GPUs to 1110 MHz \Comment{Start by capping only LP for T2}
    \Else
        \State Cap HP GPUs to 1305 MHz  \Comment{Cap HP subsequently if needed}
    \EndIf   
\ElsIf{$P > T1$}
    \State $T1cap \gets true$
    \State Cap LP to 1275 MHz \Comment{Cap LP to GPU base frequency for T1}
\EndIf   
\If{$T2cap \And P < T2-T2Buffer$}
    \State Uncap HP servers
    \State Change LP server caps to 1275 MHz
\EndIf
\If{$T1cap \And P < T1-T1Buffer$}
    \State Remove frequency cap for all LP servers
\EndIf
\EndLoop
\end{algorithmic}
\end{algorithm}

\section{Evaluation}

\begin{figure*}
    \begin{minipage}[b]{0.73\textwidth}
    \centering
    \subfloat[T1=75\%, T2=85\%.]{
        \includegraphics[width=0.30\textwidth]{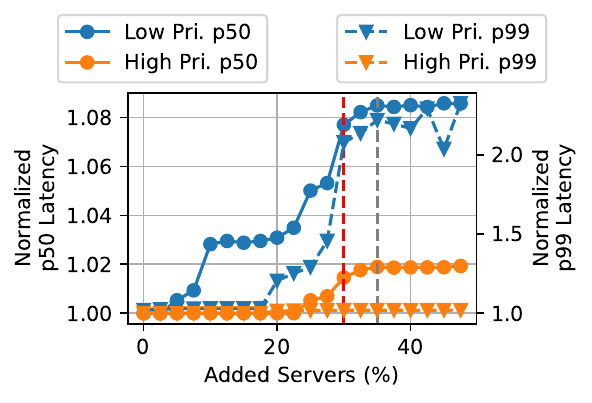}
        \label{fig:server_slowdowns_75_85}
    }
    \subfloat[T1=80\%, T2=89\%.]{
        \includegraphics[width=0.30\textwidth]{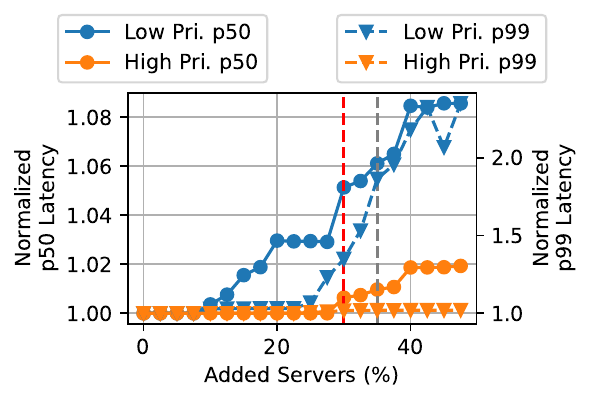}
        \label{fig:server_slowdowns_80_89}
    }
     \subfloat[T1=85\%, T2=95\%.]{
        \includegraphics[width=0.30\textwidth]{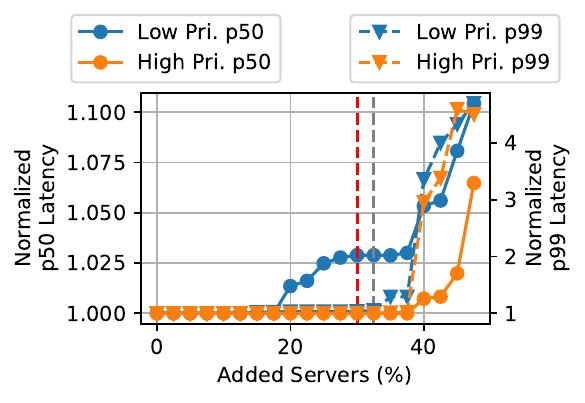}
        \label{fig:server_slowdowns_85_95}
    }   
    \caption{Threshold space search. The dashed gray line shows the max servers without powerbrake event. The dashed red line indicates adding 30\% more servers.}
    \label{fig:performance_impact_sensitivity}
    \end{minipage}
    \begin{minipage}[b]{0.23\textwidth}
    \centering
    \includegraphics[width=\textwidth]{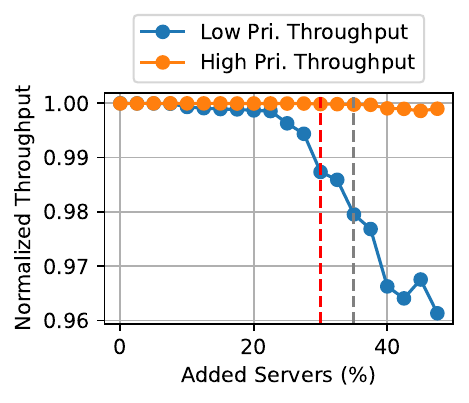}
    \caption{Server throughput for \papername.}
    \label{fig:server_throughputs}
    \end{minipage}
\end{figure*}
\label{sec:evaluation}

We first show a sweep of various parameters in the \papername policy, and then demonstrate the efficacy of \papername at oversubscribing power to allow additional capacity for LLM inference, under the defined SLOs in \Cref{tab:SLO}.

\subsection{Methodology}

We implement a discrete event simulator to evaluate the degree of oversubscription that we can support in a production LLM inference cluster.
Our simulator is optimized for a high-traffic scenario, where it assumes that all the servers in the cluster have the models loaded, and are serving inference.

\myparagraph{Workloads}
We evaluate the workloads as listed in~\Cref{tab:workload_distribution}.
We configure the BLOOM-176B model for different tasks (summarization, search, and chat) based on their input/output token sizes and priorities. Note that based on \Cref{sec:perfcharacterizeInference}, BLOOM-176B has the most performance impact from capping, making this our worst-case workload.
Each workload runs on a dedicated DGX-A100 server.

\myparagraph{Replicating production traces}
We use a six-week power consumption trace between June $21^{st}$ to August $2^{nd}$ 2023 from the production inference cluster described in \Cref{tab:cluster_power_usage}.
Based on this trace and the characteristics (\ie, power and time per token) for the open-source models, we generate a synthetic trace.
This synthetic trace contains the arrivals for each inference request, the number of tokens for the prompt and the output.
The Mean Absolute Percentage Error (MAPE) between the synthetic and original power timeseries is within 3\%.
\Cref{fig:power_utilization_timeseries} shows example traces.
Note that we only show subsets of the data and normalize the numbers in the figures for confidentiality reasons.
We use the first week to \textbf{train} the parameters of our power capping policy and \textbf{evaluate} on the subsequent five weeks.

\begin{figure}[t!]
    \centering
    \subfloat[Impact of capping frequency for low-priority workloads at T1.]{
        \includegraphics[width=0.48\columnwidth]{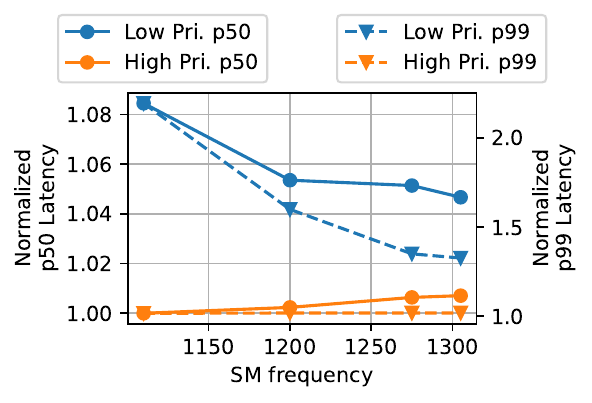}
        \label{fig:lp_capping_aggressiveness}
    }
    \subfloat[Impact of the fraction of low-priority workloads.]{
        \includegraphics[width=0.48\columnwidth]{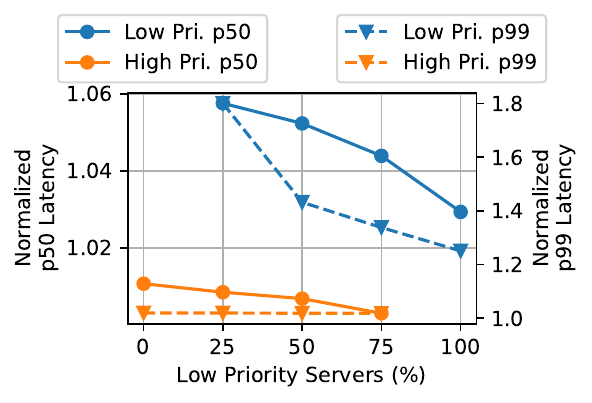}
        \label{fig:lp_vs_hp_servers}
    }
    \caption{Parameters sweeps for \papername.}
    \label{fig:fraction_lp_workloads}
\end{figure}

\subsection{Policy exploration on 1-week power trace}

\myparagraph{\papername thresholds and additional servers}
We search for the power thresholds (T1 and T2) for \papername that maximize additional servers while meeting SLOs (refer to \Cref{tab:SLO}).
We incrementally add servers, monitoring low- and high-priority latency and power brakes.
We assume the workload distribution from \Cref{tab:workload_distribution} and present a subset of results in \Cref{fig:performance_impact_sensitivity}.
T1 and T2 are set 10\% apart to ensure that capping low-priority workloads at T1 sufficiently avoids capping high-priority ones at T2.
As discussed in \Cref{sec:POLCApolicy}, based on the data in \Cref{tab:cluster_power_usage}, to allow for the maximum power spikes at 40\s granularity to be masked (11\%), our T2 should be set at 89\%.
We therefore add a T1-T2 combination of 80-89\% to the mix. 

The 75-85\% and 80-89\% T1-T2 combinations allow 35\% more servers without powerbrake (dashed gray line), while 85-95\% permits only 32.5\% more.
75-85\% misses the SLOs for low-poriority workloads by a huge margin, since it starts capping them much earlier.
On the other hand, 85-95\% incurs a lower performance impact on the LP and HP workloads, but is in a much higher danger of leading to powerbrakes, especially since the T2 is not far enough away from maximum power to avoid powerbrakes from the 40\s spikes (11\%).
To balance powerbrake avoidance and performance based on SLOs, we select $T1=80\%$ and $T2=89\%$ for \papername.
With these thresholds, we add 30\% more servers (dashed red line in \Cref{fig:performance_impact_sensitivity}) to stay strictly within the performance SLOs for LP and HP workloads.

\myparagraph{Sweeping the capping frequency}
In \Cref{fig:lp_capping_aggressiveness}, we show the performance impact of varying the capping frequency for low-priority workloads at T1.
Below 1275 MHz, we can no longer meet the SLO for LP workloads. Therefore, we choose 1275 MHz, the base frequency of A100 as the low-priority capping frequency at T1.

\subsection{Evaluation on 6-week power traces}

\begin{figure}[t!]
    \centering
    \includegraphics[width=\columnwidth]{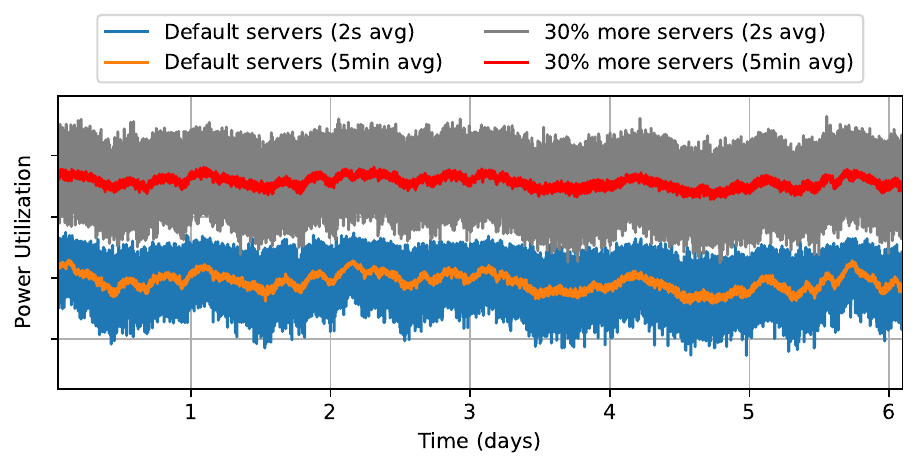}
    \caption{Row-level power utilization timeseries using BLOOM, generated based on production data.
    Y-axis hidden for confidentiality.}
    \label{fig:power_utilization_timeseries}
\end{figure}

\myparagraph{Power impact}
\Cref{fig:power_utilization_timeseries} shows the impact on daily power utilization as we add the 30\% more servers using \papername.
The main insights are: (1) the power utilization average over 5 minutes follows the same pattern with a higher power offset, and (2) the power spikes increase, since the absolute number of workloads that can be triggered together increases. 

\myparagraph{Throughput impact}
Our simulator assumes a one-request buffer per server to simulate queueing delays.
This is based on the typical load balanced setup, reducing the chance of simultaneous capping.
For the chosen thresholds, we present the impact on throughput per service for the low- and high-priority workloads.
\Cref{fig:server_throughputs} shows the high-priority workload remains unaffected, while the low-priority throughput sees a minor $<2\%$ decline.

\myparagraph{Impact of low-priority workloads}
Since we prioritize capping low-priority workloads to avoid capping high-priority workloads, the low-to-high priority distribution impacts workload performance.
\Cref{fig:fraction_lp_workloads} shows the impact on workload performance as the low- to high-priority ratio in the cluster changes.
A decrease in low-priority workloads can lead to P99 latency of high-priority workloads exceeding SLO limits.

\begin{figure}[t!]
    \centering
    \includegraphics[width=\columnwidth]{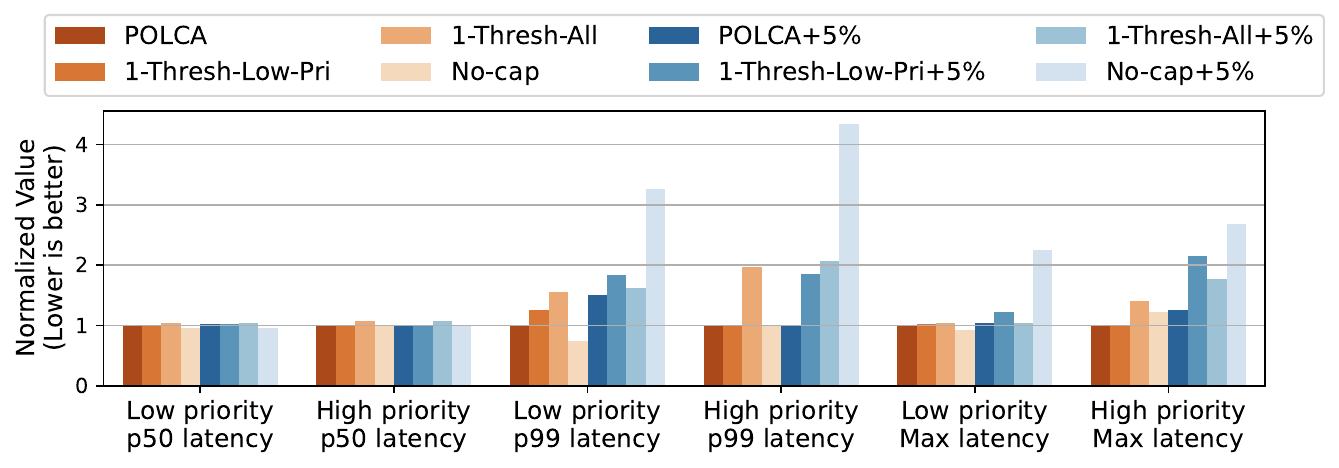}
    \caption{Performance impact of the dual-threshold  \papername with other thresholding policies at 30\% oversubscription.}
    \label{fig:power_intensive_workload}
\end{figure}

\myparagraph{Comparison with other techniques}
We compare our dual-threshold power capping policy against three baselines:
(1) single threshold at $89\%$ for low priority workloads (\texttt{1-Thresh-Low-Pri}), 
(2) single threshold at $89\%$ for all workloads (\texttt{1-Thresh-All}), and
(3) no capping (\texttt{No-cap})
All baselines include a powerbrake as fallback for power failure safety. 
The first four bars in \Cref{fig:power_intensive_workload} show the performance of various baselines normalized against \papername.

\texttt{1-Thresh-Low-Pri} does not meet the low-priority SLOs, since it directly reduces their frequency without gradual power reduction.
\texttt{1-Thresh-All} breaches the P99 SLOs for both low- and high-priority workloads, since it caps them aggressively at the $89\%$ threshold.
The \papername dual-threshold policy prioritizes the performance of high-priority workloads at the expense of the low priority workloads, still meeting SLOs for both.
\texttt{No-cap} lacks powerbrake protection, impacting P99 and P100 latency.
This policy is comparable to \papername under expected conditions, but vulnerable to model power changes.

\myparagraph{Impact of short-term changes in workloads}
We simulate the impact of workloads becoming more power-intensive than profiled.
This can happen as the same models are updated to be more efficient.
We uniformly increase the power per workload by 5\% and measure the robustness of each technique.
The last four bars in \Cref{fig:power_intensive_workload} show this performance impact.
\papername is the most robust maintaining SLOs despite the fast, inevitable workload changes.
As the changes become more prominent, and newer models take over, we reconfigure \papername.

\begin{figure}[t!]
    \centering
    \includegraphics[width=0.7\columnwidth]{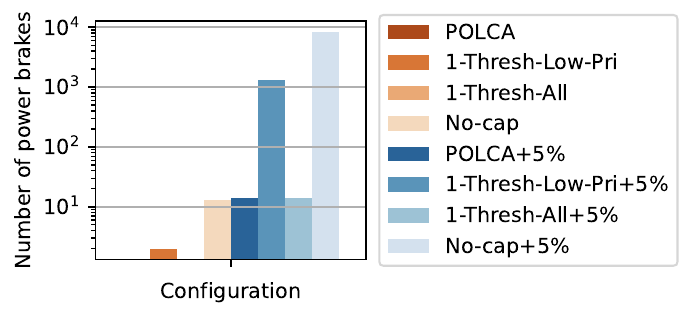}
    \caption{Number of power brakes triggered under each configuration when running default and power-intensive workloads.}
    \label{fig:power_brakes}
\end{figure}

\myparagraph{Number of powerbrake events}
Although \Cref{fig:power_intensive_workload} shows the performance impact of various policies, it is also important to track the number of powerbrake events.
The reason \papername targets and SLO of zero powerbrake events is to avoid the alarms this can cause at the cluster level for the cloud provider.
\Cref{fig:power_brakes} shows the number of powerbrake events per policy, for the regular, and scaled-up power usage workloads.

\section{Discussion}
\label{sec:discussion}

We summarize a few additional opportunities to improve and extend \papername.

\begin{figure}
    \centering
    \subfloat[Training.]{
        \includegraphics[width=0.49\columnwidth]{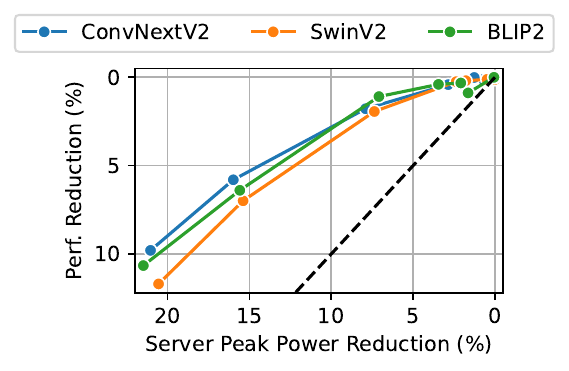}
        \label{fig:dl_training_power_vs_perf}
    }
    \subfloat[Inference.]{
        \includegraphics[width=0.49\columnwidth]{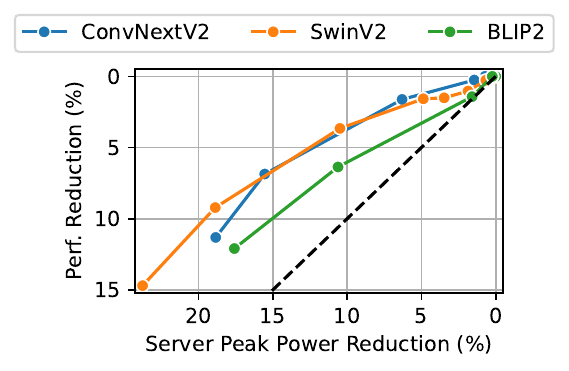}
        \label{fig:dl_inference_power_vs_perf}
    }
    \caption{Peak power reduction (relative to server TDP) vs. performance reduction for DL training and inference workloads at varying GPU SM frequencies. The dashed black line shows a linear scaling of performance with power drawn.
    }
    \label{fig:dl_power_vs_perf}
\end{figure}

\myparagraph{Extending power oversubscription insights beyond LLMs}
Unlike the prompt and token phase power patterns in generative LLMs, the vision and multi-modal deep learning inference workloads exhibit relatively stable power consumption patterns. However, they still display susceptibility to frequency scaling interventions.
\Cref{fig:dl_power_vs_perf} unveils similar response patterns in vision and multi-modal transformer models, indicating the broader applicability of power oversubscription principles.

\myparagraph{Workload-aware \papername policy} 
Given the rise of inference-as-a-service platforms, \papername could be extended to use LLM and use-case specific power profiles to reduce the impact on performance, while getting the most power savings.

\myparagraph{Using the \papername stack to mitigate training power swings}
With \Cref{fig:training_peak_power_vs_perf}, we show that the power swings during large training can be reduced at minimal performance loss using capping. The software stack that \papername provides can be tuned to handle this as per the training job size and performance profiles. 

\myparagraph{Asynchronous training}
LLM training incurs large power swings due to the synchronous iterations required across many GPUs. 
Lazy weight updates and asynchronous training could help alleviate this challenge.

\myparagraph{Phase-aware power management for generative LLMs}
Prompt phases are compute and power intensive, while token phases are not as we show in \Cref{sec:characterization}.
Customizing power-performance knobs on the GPU based on the inference phase can yield substantial savings; for example, using lower frequencies during the (longer) token phase can help reduce the average power consumption, which in turn helps free up more power to be oversubscribed.

\myparagraph{Better and standardized hardware support}
The OOB power management interfaces in GPUs today, are not only slow, but completely non-standardized. This makes building a stack that works across vendors almost impossible.
With faster, standardized out-of-band management interfaces, a lot more power and performance optimizations at scale can be achieved. In the absence of these though, there is still scope for large application owners to use in-band interfaces for better power and energy operating points.

\section{Related Work}
\label{sec:related}

Our paper is the first to extensively characterize the power patterns and power management knobs of modern LLMs in GPU clusters and propose to enable safe and efficient power oversubscription. We will discuss some of the related works.

\myparagraph{Cluster power management}
Many efforts seek to deploy the maximum number of servers possible within power capacity through power provisioning and management techniques, such as power capping and frequency scaling, focusing solely on CPUs \cite{fan2007power, ranganathan2006ensemble, pavlos2015_powercapping, li2019scalable}, power oversubscription\autocite{kumbhare2021prediction, fu2011much, govindan2009statistical}, and workload-aware placement\cite{zhang_Flex, smoothOperator}.

\myparagraph{DL energy efficiency}
Some works have looked at improving energy efficiency for both training and inference workloads through customized frameworks\cite{you2023zeus, choi2023envpipe, nie2017characterizing, nabavinejad2022coordinated} and system parameters\cite{hodak2019towards}. Reducing average power or energy consumption is different from reducing peak power, which is essential to server provisioning decisions, that \papername targets.

\myparagraph{GPU and DL workload characterization}
Many works have characterized and analyzed DL workloads in GPU clusters \cite{li2022ai, jeon2019analysis, hu2021characterization} to understand utilization and performance.
Others have studied power behaviors\cite{yu2023know, jahanshahi2020gpu} and the implications of performance under power management techniques of GPUs\cite{patki2019comparing, peres2013reverse, sinha2022not, patki2019comparing}. We are the first to extensively study the power characteristics of generative LLMs, and peak power in general.

\section{Conclusion}
\label{sec:conclusion}

We introduce \papername, a power oversubscription framework for LLM inference clusters.
With substantial characterization, we showed that generative LLM inference workloads have distinct power consumption patterns that allow for power oversubscription.
We characterized the effectiveness and limitations of existing GPU frequency scaling and power capping for LLM inference and training.
Based on our insights, we designed \papername for safe and efficient power oversubscription in LLM clusters despite unreliable and slow out-of-band GPU power management interfaces, and ever-changing models.
With production-based power simulations, we demonstrated that it can increase the allocated server capacity by 30\% in existing inference clusters, while maintaining the SLOs.
This translates to an equivalent cost and carbon reduction due to building fewer datacenters, while promptly providing much needed cluster capacity to run additional LLM workloads.

\printbibliography

\end{document}